# Time, bits, and nickel: Managing digital and analog continuity

Julie Momméja

Abstract: In 1998, the Getty Center hosted the "Time and Bits: Managing Digital Continuity" conference, gathering the founders and thinkers of two San Francisco non-profit organizations interested in long-term thinking and archiving: the Internet Archive and the Long Now Foundation. This chapter proposes to discuss two different ways of archiving through time, in digital and analog formats, for virtual web contents and physical paper-based ones. It explores various types of archiving methods and tools and the management challenges they raise, in terms of time and space, but also innovation, maintenance, and "continuity". It depicts two distinct visions of the future of archiving which nonetheless converge in their mission of safeguarding, sharing, and giving access to information and knowledge for the decades and centuries to come..

Introduction

*"One of the peculiar things about the 'Net is it has no memory. (…) We've made our digital bet. Civilization now happens digitally. And it has no memory. This is no way to run a civilization. And the Web—its reach is great, but its depth is shallower than any other medium probably we've ever lived with." (Kahle 1998)*

In February 1998, the Getty Center in Los Angeles hosted the Time and Bits: Managing Digital Continuity conference, organized by the founders and thinkers of two non-profit organizations established two years earlier in San Francisco: the Internet Archive and the Long Now Foundation, both dedicated to long-term thinking and archiving.

The Internet Archive has since become a global example of digital archiving and an open library that provides access to millions of digitized pages from the web and paper books on its website. The Long Now Foundation's central project involves the design and construction of a monumental clock intended to tick for the next 10,000 years, promoting long-term thinking alongside its lesser-known archival mission: the Rosetta Project.

The Time and Bits: Managing Digital Continuity conference brought together these organizations to discuss what the Internet Archive's founder, Brewster Kahle, referred to as "our digital bet": a digital-only novel form of civilization with no history, posing challenges to the preservation of its immaterial cultural memory. Discussions at the conference raised concerns about the longevity of digital formats and explored potential archival and

Julie Momméja, University of Lorraine, France, julie.mommeja@univ-lorraine.fr, 0000-0003-1148-2490

transmission solutions for the future. This foresight considered the 'heritage' characteristics of the digital world, which were yet to be defined as such on a global level. It was not until 2003 that UNESCO published a charter advocating for "digital heritage conservation", distinguishing between "digital-born heritage" and digitized heritage (UNESCO 2003, Musiani et al. 2019). The Getty Center conference thus emerges as a precursor in the quest to preserve both digital data and analog information for future generations. This endeavor, the chapter argues, aligns with the *longue durée*, a conception of time and history developed by French historian Fernand Braudel during World War Two (Braudel 1958).

While the Internet Archive has envisioned such a mission through the continuous digital recording of web pages and the digitization of paper, sound, and video documents into bits format, the Long Now Foundation's Rosetta Project began to take shape during the 'Time and Bits' conversations, offering a different approach to data conservation in an analog microscopic format, engraved on nickel disks.

Taking the 1998 gathering of the Internet Archive and Long Now Foundation as a starting point, this chapter aims to examine the challenges and strategies of 'digital continuity management' (or maintenance). It proposes to analyze the different ways these two case studies envision archiving and transmission to future generations, in both digital and analog formats—bits and nickel, respectively—for virtual web content and physical paper-based materials.

Through a comparative analysis of these two non-profit organizations, this chapter seeks to explore various archiving methods and tools, and the challenges they present in terms of time, space, innovation, maintenance, and 'continuity'. By depicting two distinct visions of the future of archiving represented by these organizations, it highlights their shared mission of safeguarding, sharing, and providing universal access to information, despite their differing formats.

The method used for this analysis combines theoretical, comparative, and qualitative studies through an immersive research process spanning over three years in the San Francisco Bay Area. This process involved conducting interviews and participant observations during seminars, talks, and meetings held by both the Long Now Foundation and at the Internet Archive. Additionally, archival research was conducted online, using resources such as the Internet Archive's Wayback Machine and the Long Now Foundation's blog dating back to 1996, as well as on-site at the Long Now Foundation.

This chapter adopts a multidisciplinary approach conjoining history, media, maintenance, and American studies to analyze the challenges faced by these two organizations in transmitting both material and intangible cultural heritage (UNESCO 2003).

The first part of this chapter concentrates on the future of archives and their longevity, a topic that was discussed during the 1998 Time and Bits conference. It suggests a parallel with the Braudelian *longue durée* perspective, which offers a novel understanding of time and history.

The second part focuses on digital transmission in 'hard-drive form' using the example of the Internet Archive and its Wayback Machine, comprising thousands of hard drives. The final segment of this chapter discusses the analog archival format chosen by the Long Now Foundation, represented by the Rosetta Disk, a small nickel disk engraved with thousands of pages of selected texts. This format is likened to a modern iteration of the Egyptian Rosetta Stone for preservation into the long-term future.

1. "Time and Bits: Managing Digital Continuity"…and maintenance in *longue durée*

"How long can a digital document remain intelligible in an archive?" This question, asked by futurist Jaron Lanier in one of the hundreds of messages posted on the Time and Bits forum that ran from October 1997 until June 1998, underscores not only concerns about the future 'life' of digital documents at the end of the 1990s, but also their meaning and understanding in archives for future generations. These concerns about digital preservation were central to discussions at the subsequent Time and Bits conference organized a few months later in February 1998 at the Getty Center in Los Angeles by the Getty Conservation Institute and the now-defunct Getty Information Institute, in collaboration with the Long Now Foundation.

The Long Now Foundation, formed in 1996, emerged from discussions among thinkers and futurists who later became its board members This group included Stewart Brand, recipient of the 1971 National Book Award for his *Whole Earth Catalog* and co-founder of the pioneering virtual community, the WELL, created in the 1980s (Turner 2006), engineer Danny Hillis, British musician and artist Brian Eno, technologists Esther Dyson and Kevin Kelly, and futurist Peter Schwartz (Momméja 2021). Schwartz, in particular, is the one who articulated the concept of the 'long now' as a span of 20,000 years—10,000 years deep into the past and 10,000 years into the very distant future. This timeframe coincides with the envisioned lifespan of the monumental Clock being constructed by the foundation in West Texas. The choice of this specific duration marks the end of the last ice age about 10,000 years ago, a period that catalyzed the advent of agriculture and human civilization, with some scholars even identifying it as the onset of the Anthropocene epoch. Indeed, a group of scientists extends their analysis beyond the industrial era, which has generally been studied as the beginning of this human-induced transformation of our biosphere, considering the origins of agriculture as "the time when large-scale transformation of land use and human-induced species and ecosystem loss extended the period of warming after the end of the Pleistocene" (Henke and Sims 2020). For the founders of the Long Now Foundation, this 10,000-year perspective must therefore be developed in the opposite direction, towards the future (hence the expected duration of the Clock) forming the 'Long Now'.

The paper argues that 'long now' promoted by the organization can be paralleled with the concept of *longue durée* put forth by Annales historian Fernand Braudel. Braudel began elaborating the idea of *longue durée* during his time as prisoner of war in Germany. For five years, he diligently worked on his PhD dissertation, *La Méditerranée et le monde méditerranéen à l'époque de Philippe II* (Braudel 1949). It was during his internment that Braudel developed the concept of the 'very long time', a temporal construction that provided him solace from the traumatic events he experienced in the 'short time' and helped him gain insight into his condition by situating them on a much broader time scale (Braudel 1958). With newly stratified temporalities—from the immediate to the medium to the very long term—Braudel succeeded in escaping the space-time of which he was a prisoner, a 'here' and 'now' devoid of meaningful perspectives when a longer 'now' would liberate him from the present moment. *Longue durée* was thus imagined as a novel long-term approach to history, diverging from traditional narratives that focused on brief periods and dramatic events, such as wars. This is what Braudel referred to as "a rushed, dramatic narrative" (Braudel 1958). A second, longer type of history, based on economic cycles and conjunctures, was described by Braudel as spanning several decades, while *longue durée* offered a novel type of history that transcended events and cycles, extending even further to encompass centuries—although the French historian refrained from specifying an exact timeframe.

*Longue durée*, alongside its modern Californian counterpart, the 'long now', prompts us to reconsider our understanding of history in time as a means to encapsulate events far beyond our lifetimes. Braudel insisted historians should incorporate *longue durée* into their work and rethink history as an 'infrastructure' composed of layers of 'slow history'.

Given this perspective, how can we archive and transmit fast traditional history within the context of *longue durée*? In his foreword to the Time & Bits report, Barry Munitz, president and CEO of the J. Paul Getty Trust, explained the initiative behind the conference:

> We take seriously the notion of long-term responsibility in the protection of important cultural information, which in many cases now is recorded only in digital formats. The technology that enables digital conversion and access is a marvel that is evolving at lightning speed. Lagging far behind, however, are the means by which the digital legacy will be preserved over the long term (Munitz 1998).

The two organizations selected for this chapter offer two distinct, yet complementary, visions of how archiving and transmitting should be approached, now and for the *longue durée*, in digital and analog formats.

2. Digital transmission in 'hard-drive form': The Wayback Machine and the Internet Archive

Addressing the "problem of our vanishing memory" was a focal point of the Time & Bits conference encapsulated by Internet Archive founder Brewster Kahle's question: "I think the issue that we are grappling with here is now that our cultural artifacts are in digital form, what happens?" (Kahle 1998). As noted by Stewart Brand, Kahle also pointed out that "one of the peculiar things about the 'Net is it has no memory. (…) We've made our digital bet. Civilization now happens digitally. And it has no memory. This is no way to run a civilization. And the Web—its reach is great, but its depth is shallower than any other medium probably we've ever lived with" (Kahle 1998).

As a way to resolve this 'digital bet' and the pressing need for 'digital continuity', Brewster Kahle embarked on a mission to archive the web on a massive scale, giving rise to the Internet Archive and its Wayback Machine: an archive comprising 20,000 hard drives and containing 866 billion web pages as of March 2024.

Like the Long Now Foundation, the Internet Archive is a non-profit organization founded in 1996 in San Francisco. In fact, both entities once occupied adjacent offices in the Presidio. Their missions can also be put in parallel: whereas the Long Now Foundation promotes long-term thinking through projects like the construction of a Clock and the preservation of foundational languages and texts of our civilization in analog form through the Rosetta Disk, the Internet Archive digitizes and archives analog documents and records digital textual heritage through its Wayback Machine.

The Internet Archive embarked on its mission with an imperative to save internet pages, immaterial data composed of bits, which had not previously been archived: "We began in 1996 by archiving the internet itself, a medium that was just beginning to grow in use. Like newspapers, the content published on the web was ephemeral—but unlike newspapers, no one was saving it" (Internet Archive 2024). Despite the transient and intangible nature of web pages, the Internet Archive remains committed to this mission, continuing to archive internet pages in a digital format to this day, with the ambition to remain open and collaborative, "explicitly promoting bottom-up initiatives intended to revalue human intervention" (Musiani et al. 2019).

Brewster Kahle, who could be regarded as the first digital librarian in history, promotes "Universal Access to All Knowledge" and "Building Libraries Together". These missions, as explained during the Internet Archive's annual celebration on October 21, 2015, at its headquarters in San Francisco, highlight the organization's commitment to a wide array of digital content, including internet pages, books, videos, music, and games. Therefore, the internet appears as a "heritage and museographic object" (Schafer 2012), with information worth saving and protecting for the future. While the Library of Congress recently acknowledged the significance of Twitter content as a form of heritage (Schafer 2012), the Internet Archive has been standing as an advocate for the preservation and transmission of digital heritage as early as the 1990s. UNESCO further validated this recognition in 2003 by acknowledging the existence of "digital heritage as a common heritage" through a charter on the conservation of digital heritage (Musiani et al. 2019) where resources are 'born digital', before being, or even without ever being, analog:

> Digital materials encompass a vast and growing range of formats, including texts, databases, still and moving images, audio, graphics, software, and web pages. Often ephemeral in nature, they require purposeful production, maintenance, and management to be retained. Many of these resources possess lasting value and significance, constituting a heritage that merits protection and preservation for current and future generations. This ever-growing heritage may exist in any language, in any part of the world, and in any area of human knowledge or expression (UNESCO 2003).

The Internet Archive's mission aligns perfectly with this definition, providing open access to documents that are "protected and preserved for current and future generations", echoing once again the Long Now Foundation's own mission. However, the pursuit of "universal access to all knowledge" raises questions about the quality or "representativeness of the archive" (Musiani et al. 2019) in the face of the abundance and diversity of the sources and formats available.

For instance, the music section of the Internet Archive connects visitors to San Francisco's local counterculture history with a vast collection of recordings from Grateful Dead shows (17,453 items) that fans contributed to the organization in analog formats for digitization. This exchange has not only allowed the band's fan community to flourish but has also bolstered the group's the popularity: "they started to record all those concerts and you know, there are I think 2,339 concerts that got played by the Grateful Dead (…) and all but 300 of those are here in the archive" (Barlow 2015). In this way, the Internet Archive confirms its role as a universal collaborative platform and effectively contributes to a "new era of cultural participation" (Severo and Thuillas 2020), one that is proper to Web 2.0 but which the non-profit has been championing since the 1990s.

However, for the Internet Archive, and digital technology in general, to truly guarantee the archiving of human heritage 'for future generations' over the years, whether initially analog or digital, it is imperative to continuously improve and update storage formats and units to combat obsolescence and adapt to evolving technologies:

> Of course, disk drives all eventually fail. So we have an active team that monitors drive health and replaces drives showing early signs for failure. We replaced 2,453 drives in 2015, and 1,963 year-to-date 2016… an average of 6.7 drives per day. Across all drives in the cluster the average 'age' (arithmetic mean of the time in-service) is 779 days. The median age is 730 days, and the most tenured drive in our cluster has been in continuous use for 6.85 years! (Gonzalez 2016)

If "all contributions produced on these platforms, whether amateur or professional, participate in the construction and appropriation of cultural and memorial heritage" (Severo and Thuillas 2020), reliance solely on digital technology poses a substantial challenge to the preservation of our cultures in *longue durée*. Aware of the inherent risks associated with archiving both analog and 'digital heritage' on storage mediums with limited lifespans, the Internet Archive must make the maintenance and replacement of the hard drives that comprise its Wayback Machine a constant priority.

3. From stone to disk: the Rosetta Project through time and space

To embody Braudel's notion of 'slow history' and foster long-term thinking among people, the Long Now Foundation envisioned not only a monumental Clock as a time relay for future generations, but also a library for the deep future, soon materializing as an engraved artifact: The Rosetta Disk.

As explained by technologist Kevin Kelly, the concept of a miniature storage system comprising 350,000 pages of text engraved on a nickel disk, measuring just under eight centimeters in diameter, was proposed by Kahle during the Time and Bits: Managing Digital Continuity conference, "as a solution for long-term digital storage (…) with an estimated lifespan of 2,000–10,000 years" (Kelly 2008). These meeting discussions thus led to the emergence of the Rosetta Project within the Long Now Foundation, drawing inspiration from the Rosetta Stone. The final version of the Rosetta Project's Disk was unveiled in 2008: 14,000 pages of information in 1,500 different languages (Welcher 2008). Crafted in analog format, it was conceived as the solution to the ever-changing landscape of digital technologies.

While the Internet Archive possesses infinite possibilities for archiving, the Long Now Foundation's analog choice demands a thoughtful selection of texts to be micro-engraved onto the disk. The foundation decided to focus on several texts, both symbolic and universalist, such as the 1948 Universal Declaration of Human Rights, along with Genesis, chosen for its numerous translations. Materials with a linguistic or grammatical vocation, such as the Swadesh list—a compendium of words establishing a basic lexicon for each language—were included, as well as grammatical information including descriptions of phonetics, word formation, and broader linguistic structures like sentences.

Unlike Kahle's digital and digitized heritage project, the Foundation's language archive is exclusively engraved, accessible only through a microscope. Such an archive is thus a finite heritage, with no scope for future development beyond the creation of new disks displaying new texts. While the Internet Archive and its Wayback Machine are constantly evolving, updated through constant digitization and the preservation of new web pages, the format and size of the nickel disk remain immutable.

To ensure the long-term survival of this archive, the foundation has embraced the "LOCKS" principle—Lots of Copies Keep Stuff Safe—and has opted to duplicate its Rosetta Disk. By distributing these duplicates worldwide, the project stands a greater chance of lasting in *longue durée*: "this project in long-term thinking would do two things: it would showcase this new long-term storage technology, and it would give the world a minimal backup of human languages" (Kelly 2008).

The final version of the Rosetta Disk, containing 14,000 micro-engraved pages, was presented at the Foundation's headquarters in 2008. "Kept in its protective sphere to avoid scratches, it could easily last and be read 2,000 years into the future" (Welcher 2008). Beyond its resilience within the timeline of the Long Now, the analog Rosetta Disk aspires to endure across space as well. Remarkably, as the Foundation had been developing its project since 1999, they were contacted by the European Space Agency (ESA) and the Rosetta Mission team which, coincidentally, was working on the launch of an exploratory space probe aptly named Rosetta. The Rosetta probe was launched on March 2, 2004, aboard an Ariane 5G+ rocket from Kourou, with the mission of studying comet 67P/Churyumov-Gerasimenko ('Tchouri') located near Jupiter. On board the probe was the very first version of the Rosetta Disk, less comprehensive than the version unveiled in 2008, nevertheless containing six thousand pages of translated texts.

Conclusion

On November 12, 2014, over a decade after its departure from Earth, the Rosetta probe finally reached Comet Tchouri. Upon arrival, it deployed its Philae lander onto the comet's surface, where, despite unexpected rebounds, it eventually stabilized itself to conduct programmed analyses. Nearly two years later, on September 30, 2016, the Rosetta module, with the Rosetta Disk on board, joined Philae on Tchouri, thus marking the conclusion of the mission: "With Rosetta we are opening a door to the origin of planet Earth and fostering a better understanding of our future. ESA and its Rosetta mission partners have achieved something extraordinary today" (ESA 2014). Through a space mission focused on the future with the aim of better understanding the Earth's past, the Rosetta Disk fulfilled its project to become an archive in *longue durée*, transcending temporal and spatial boundaries.

Almost ten years later, both the Rosetta Disk and the Internet Archive, through a selection of books and documents from its datasets, became part of an even larger spatial archive which also includes articles from Wikipedia and books from Project Gutenberg, all etched on thin sheets of nickel. The Arch Mission Foundation's Lunar Library successfully landed on the Moon on February 22, 2024, thus reuniting for the first time the two non-profits' archival materials in a cultural and civilizational preservation project, built to remain on the Moon surface throughout the *longue durée*.

The Time and Bits: Managing Digital Continuity conference did not present a single solution to the challenges of digital archives and data transmission. Instead, it offered a range of options and tools for web archives, digital data, and analog documents to address our 'digital bet'. The two cases presented appear as two faces of the same disk—digital and analog—with a shared conservation objective: providing different means to consider *longue durée* and ensure archival continuity and maintenance in the long term. This continuity extends not only through time, but also across space, placing "digitally-born heritage" (Musiani et al. 2019) and more traditional forms of heritage on equal footing.

From the "creative city" (Florida 2002) of San Francisco, both organizations have managed to extend the boundaries of the "creative Frontier" (Momméja 2001), not only physically and digitally, but also through *longue durée* and space. From hard drives to disks, they offer a new form of coevolution between humans and machines, a 'post-coevolution' aimed at transmitting our cultural heritage to future generations through bits and nickel.